\documentclass{emulateapj}

\newcommand{\msol}{\hbox{${\rm M}_\odot$}}

\newcommand{\kms}{km s$^{-1}$}
\newcommand{\hinv}{h^{-1}}
\newcommand{\mpc}{\rm{Mpc}}
\newcommand{\hmpc}{\hinv\mpc}
\newcommand{\kpc}{\rm{kpc}}
\newcommand{\hkpc}{\hinv\kpc}

\shorttitle{Satellites in MW-like hosts}
\shortauthors{Gonz\'alez, Kravtsov \& Gnedin}

\begin{document}
\title{Satellites in MW-like hosts: Environment dependence and close pairs.}

\author{Roberto E. Gonz\'alez\altaffilmark{1,2}}
\author{Andrey V. Kravtsov\altaffilmark{1,2,3}}
\author{Nickolay Y. Gnedin\altaffilmark{4,1,2}}
\altaffiltext{1}{Department of Astronomy \& Astrophysics, The University of Chicago, Chicago, IL 60637 USA {\tt regonzar@oddjob.uchicago.edu}}
\altaffiltext{2}{Kavli Institute for Cosmological Physics, The University of Chicago, Chicago, IL 60637 USA}
\altaffiltext{3}{Enrico Fermi Institute, The University of Chicago, Chicago, IL 60637 USA}
\altaffiltext{4}{Particle Astrophysics Center, Fermi National Accelerator Laboratory, Batavia, IL 60510 USA}

\begin{abstract}

Previous studies showed that an estimate of the likelihood distribution of the Milky Way halo mass can be derived using the properties of the satellites similar to the Large and Small Magellanic Clouds (LMC and SMC). However, it would be straightforward to interpret such an estimate only if the properties of the Magellanic Clouds (MCs) are fairly typical and are not biased by the environment.  In this study we explore whether the environment of the Milky Way affects the properties of the SMC and LMC such as their velocities. To test for the effect of the environment, we compare velocity distributions for MC-sized subhalos around Milky Way hosts in a sample selected simply by mass and in the second sample of such halos selected with additional restrictions on the distance to the nearest cluster and the local galaxy density, designed to mimic the environment of the Local Group (LG). We find that satellites in halos in the LG-like environments do have somewhat larger velocities, as compared to the halos of similar mass in the sample without environmental constraints. For example, the fraction of subhalos matching the velocity of the LMC is $23\pm2\%$ larger in the LG-like environments.  
We derive the host halo likelihood distribution for the samples in the LG-like envirionment and in the control sample and find that the environment does not significantly affect the derived likelihood. We use the updated properties of the SMC and LMC to derive the constraint on the MW halo mass $\log{({\rm M}_{200} /\msol)}=12.06^{+0.31}_{-0.19}$ ($90\%$ confidence interval). We also explore the incidence of close pairs with relative velocities and separations similar to those of the LMC and SMC and find that such pairs are quite rare among $\Lambda$CDM halos. Only $2\%$ of halos in the MW mass range have a relatively close pair (\mbox{$\Delta r<40\kpc$} and \mbox{$\Delta s<160$ \kms}) of subhalos with circular velocities \mbox{$v_{\rm circ}>50$ \kms}. Pairs with masses and separations similar to those of the LMC and SMC ($\Delta r_{\rm MC}=23.4\pm10$ \kpc, and $\Delta s_{\rm MC} =128\pm32$ \kms) are found only in one out of $\approx 30000$ MW-sized halos. Interestingly, the halo mass likelihood distribution for host halos constrained to have MC-like close pairs of subhalos  is quite different from the global likelihood from which the MW halo mass constraint discussed above was derived. Taking into account the close separation of the MCs in the Busha et al.\ 2011 method results in the shift of the MW halo mass estimate to smaller masses, with the peak shifting approximately by a factor of two.

\end{abstract}


\section{Introduction}


In the CDM scenario the Milky Way (MW) halo formed by accretion and disruption of smaller halos, and some of them survived this process as self-bound substructures or subhalos orbiting around the host halo \citep[e.g.,][see \citealt{2010AdAst2010E...8K} for a recent review]{1999ApJ...522...82K, 1999ApJ...524L..19M}.
The majority of these subhalos are likely devoid of stars, but some of them are massive enough to host dwarf galaxies \citep[e.g.,][]{kauffmann_etal93,1999ApJ...522...82K,bullocketal00,kravtsov_etal04,2011MNRAS.417.1260F}.
The MW and its satellites offer a unique laboratory for testing CDM predictions for halo formation and associated substructure, because properties of many of the satellites have been studied in details in observations, including proper motions and rotation curves \citep[e.g.,][]{2006ApJ...638..772K,2006ApJ...652.1213K}. 

It is still debated how typical is the satellite population of the MW, but both theoretical models and observations indicate that the incidence of the Magellanic Clouds (MCs) satellites is rare for MW-sized galaxies.
Semi-analytic galaxy formation models predict that only $\simeq 10\%$ of the MW-sized galaxies have satellites as bright as the LMC \citep[e.g.,][]{1993MNRAS.264..201K,2009ApJ...696.2179K}. \citet{2011MNRAS.414.1560B} used large sample of MW-sized halos extracted from the Millenium-II simulation and found that $20\%$ MW-mass halos host an LMC or SMC, and only $\sim 2.5 \%$ host both MCs. A similar frequency was found by \citet{2011ApJ...743..117B} using the Bolshoi simulation of the concordance cosmology and abundance matching ansatz to assign stellar masses to halos and subhalos.
In observations,
\citet{2011MNRAS.411..495J} searched for star-forming satellites around $143$ luminous spiral galaxies and found that two-thirds of central galaxies have no satellites down to luminosity and star formation rates well below those of the MCs. Using the Sloan Digital Sky Survey (SDSS), \citet{2011ApJ...733...62L} computed the occurrence of the satellites similar to the MCs in luminosity around MW-sized galaxies and found that $11\%$ have one, and only $3.5\%$ have two MCs within a radius of $150$\kpc.
Similar values for the frequency of MCs are found in the SDSS in a recent study by \citet{2011ApJ...738..102T}, and in the GAMA survey by  \citet{2012MNRAS.424.1448R}. What makes Magellanic Clouds even more peculiar is the fact that they are likely to be an interacting pair (see \S~\ref{sec:closepairs} below) and have rather high velocities with respect to the Milky Way. 

One caveat in this debate is that some satellite properties depend sensitively on the MW halo mass \citep[e.g.,][]{wang_etal12}.  
Although there are several methods to constraint the MW halo mass \citep[see e.g.,][and references therein]{2011ApJ...743...40B,2012MNRAS.424L..44D}, it is still quite uncertain. In this study we focus on the methods that use satellite properties to constrain the MW halo mass.
In particular, as shown by \citet[B11 hereafter]{2011ApJ...743...40B}, the MW host halo mass likelihood can be computed using the observed MCs properties and a statistically representative sample of CDM halos to evaluate the likelihood that a given halo would have each or all MCs properties, such as circular velocity, host halo distance, and velocity within the host. Such approach provides a new interesting constraint on the MW virial mass, but there are certain questions that need to be addressed to assess its reliability and interpretation. One of such questions is whether the environment of the Local Group biases properties of the MW satellite population, and MCs in particular. We explore this and other related questions in more detail in this paper.

Another important question is the origin and dynamics of the MCs.
The accretion history of the MCs is still not clear, the presence of the Magellanic Stream (MS hereafter), a filament of gas extending $150^{\circ}$ across the sky, with an apparent spatial and chemical association with the MCs, is interpreted as a tidal tail \citep[see][and references therein]{besla_etal12}.
There are several clues which indicate that the MCs are bound, and have been interacting recently \citep{2012ApJ...750...36D,2012MNRAS.421.2109B}, and there is also some evidence of the SMC stars accreted onto the LMC \citep{2011ApJ...737...29O,2012ApJ...753..123C}. 
Close satellite pairs are rare in nearby MW-sized galaxies \citep{2012MNRAS.424.1448R,2011MNRAS.411..495J}.
Proper motion measurements for the MCs \citep{2006ApJ...638..772K,2006ApJ...652.1213K} indicate that they have high velocities not aligned with the Magellanic Stream, and suggest that the MCs could be bound and on their first or second pericenter passage into the MW \citep{2007ApJ...668..949B,2010ApJ...721L..97B,2012MNRAS.421.2109B}.
Other studies also suggest that the MCs were accreted in the same system \citep{2008ApJ...686L..61D,2011ApJ...742..110N}. B11 estimated from the Bolshoi simulation that, for a typical MW-sized host, there is a $\simeq 72\%$ probability that the MCs were accreted within the last Gyr. \citet{2011MNRAS.414.1560B} and \citet{besla_etal12} also favor the late accretion scenario for the MCs.

The paper is organized as follows.
We describe the simulation and halo catalogs used in our study in \S~\ref{sec:sims}, while in \S~\ref{sec:data} we describe the data samples for our \mbox{MW-}, \mbox{MCs-}, and \mbox{LG-analogues}.
We presents the MW halo mass likelihood in different environments and under different assumptions on whether MCs are independent velocity samples in \S~\ref{sec:results} and \S~\ref{sec:closepairs}. We present discussion and our conclusions in \S~\ref{sec:conclusions}.

\section{Simulation and halo catalogues}
\label{sec:sims}

To carry out our analysis we use halos extracted from the Bolshoi simulation \citep{2011ApJ...740..102K}, which followed evolution of $2048^3$ particles in $250 \hmpc$ cubic volume assuming concordance flat $\Lambda$CDM cosmology with parameters: $\Omega_{\rm m}=1-\Omega_\Lambda=0.27$, $\Omega_{\rm b}=0.0469$, $h=H_0/(100)=0.7$, $\sigma_8=0.82$ and $n_{\rm s}=0.95$, compatible with combined constraints from WMAP, BAO, SNe, and cluster abundance \citep{komatsu_etal11}. The high spatial and mass resolution and relatively large volume make the Bolshoi simulation well suited for providing a base halo sample for our study.

 The halo catalog we used was constructed using the Bound Density Maxima halo finder \citep{1997astro.ph.12217K,klypin_etal99a}, which identified $2285977$ halos down to the resolution completeness limit of $v_{\rm circ}\approx50$ \kms (corresponding to $\approx 110$ particles or $M_{200} \approx 1.7 \times 10^{10}$ \msol). The circular velocity $v_{\rm circ}$ is the maximum of the circular velocity profile $v_{\rm circ}(r)=\sqrt{GM(<r)/r}$. The parameters of the halo finder were set such that the density maxima are not allowed to be closer than $10 \hkpc$, and the finder keeps only the most massive density maximum if that happens. Halo center is identified with the particle location which has the largest local density, and bulk halo velocity is computed as the average velocity of the $30$ closest neighbors of the central particle. The algorithm computes a number of halo properties after an iterative procedure to remove unbound particles.

Throughout this paper we defined halo mass, $M_{200}$, as the mass enclosed in a sphere of radius $R_{200}$ with the density $200$ times the critical density of the universe at the redshift of analysis. Another common mass definition is $M_{vir}$, the mass within the radius enclosing the mean overdensity of 358 with respect to the mean density of the universe (or overdensity of $358\times 0.27\approx 97$ with respect to the critical density) \citep{1998ApJ...495...80B}. 
In the Bolshoi simulation, we find a relation of $M_{\rm vir}/M_{200}=1.21$ for host halos of $v_{\rm circ}\approx 220$ \kms.

We use MW-sized host halos in the Bolshoi catalogs to search for subhalos with velocities and positions similar to those of the MCs. However,  positions of satellites change on short timescales due to their motion along their orbits. Therefore, to increase statistics, we stack the halo catalogs of several simulation snapshots close to $z=0$ separated by $\Delta a=0.003$ \mbox{($\sim 42 \rm{Myrs}$ at $z=0$)}. During the time interval between snapshots a typical MC should move $\sim 15 \kpc$ along its trajectory. We stack satellites of the last $30$ snapshots, and the total time difference between the first and last snapshot is $\sim 1.3 \rm{Gyrs}$ (or $\Delta z < 0.1$), so we can neglect any evolution effect in MW mass halos \citep{2011MNRAS.411..584M,2012arXiv1207.0816D,2008MNRAS.389..385C}.

For a given MW-sized halo at $z=0$ we should, in principle, consider the last $N$ simulation outputs to trace trajectories of all satellites and check if they match MCs contraints at some time in the recent past. However, this is computationally expensive. Instead,  in our analysis we consider each snapshot as an independent realization of halo properties, which effectively increases the simulation sample by a factor of $N$.  In this approximation we neglect any correlation in the positions and velocities of satellites between snapshots. For example, in the extreme case of a purely circular orbit,  a single LMC or SMC analogue would be counted $N$ times. However, for realistic eccentric orbits such double counting is quite rare, especially for satellites with relatively small radial distances to host centers similar to those of the SMC and LMC.  
On the other hand, we have a large number of hosts and satellites in each snapshot, and the randomness of their orbital configuration produces consistent distributions and average fraction of satellites matching any given set of constraints, for any number of snapshot selected. We have tested for the double counting effects using different number of snapshots, $N$, computing the distribution of satellites matching several set of constraints, and we have found no significant differences in the properties of satellites for $N\la30$.

\section{Halo samples and observational constraints}
\label{sec:data}

In our analysis we use the following three main observational measurements.  

\subsection{The Milky Way halo mass}

To select MW analogues from the Bolshoi-derived halo catalogues, we select halos within the broad mass range $M_{200c}=0.8 - 2.9 \times 10^{12}$ \msol, which covers the range of current observational constraints: e.g., using HI gas distribution \citep{2007AA...469..511K}, 
kinematics of stars \citep{2009PASJ...61..227S,2008ApJ...684.1143X,2010ApJ...720L.108G,1999MNRAS.310..645W,2012MNRAS.424L..44D}, satellite dynamics \citep{2010MNRAS.406..264W}, escape velocity \citep{2007MNRAS.379..755S}, and timing argument \citep{2008MNRAS.384.1459L}.
There are ($\sim 57000$) host halos in this mass range in the Bolshoi simulation at $z=0$, which contain $\sim 115000$ subhalos with $v_{\rm circ}>50$ \kms.
 
\subsection{Magellanic Cloud analogues}

We follow \citet{2011ApJ...743...40B} and use the following observed properties of the Magellanic Clouds to select appropriate MC analogues among subhalos in the MW halo analogues and to constrain the halo mass of the Milky Way: the distance to the host center $r_0$, the total speed relative to the host center $s$, and the subhalo circular velocity $v_{\rm circ}$. For these quantities we use recent HST measurements by \citet{kal2012}: $v_{\rm circ}=76.1 \pm 7.6$ \kms, $r_0=50 \pm 5 \kpc$, $s=321 \pm 24$ \kms for the LMC, and $v_{\rm circ}=60 \pm 5$ \kms, $r_0=60 \pm 5 \kpc$, $s=217 \pm 26$ \kms for the SMC. 
The halos hosting the MCs analogues are required to have at least two subhalos with $v_{\rm circ}>50$ \kms and we will consider only the two subhalos with the largest circular velocities\footnote{In B11, the MCs analogues are selected in hosts that have exactly two subhalos with $v_{\rm circ}>50$ \kms, we will explore the effects of this difference in the Results section.}.

Note that $r_0$ errors are inflated from their actual observational values to improve statistics of the sample of the MC analogues. The range of $r_0$ we use corresponds to typical radial displacement of subhalos along its orbit between consecutive snapshots.
These parameters are used everywhere in this paper, except in \S~4.2, where we use a different definition for the MCs analogues assuming a single bound system, and in \S~5, where we include two additional constraints: the relative separation and velocity of the clouds.

\subsection{The Local Group analogues}
\label{sec:LGanalogs}

The Milky Way is not an isolated galaxy, but is located in a pair with M31 \citep[$\approx 770$~kpc away][]{2005MNRAS.356..979M,2005ApJ...635L..37R} and is surrounded by a number of smaller galaxies, collectively known as the Local Group of galaxies. On larger scales the environment of the Local Group is rather low-density: in a sphere of $50\ \mpc$ radius around the LG, the estimated density is $\sim 3$ times lower than average \citep{karachenstev}, while in a sphere of $5\ \mpc$ the density around the Local Group is close to the mean density of the universe \citep{klypin_etal03,2005AJ....129..178K}. In addition, the nearest cluster to the Local Group is the Virgo Cluster $\sim 16.5\ \mpc$ away \citep{2007ApJ...655..144M}. 

To explore whether environment of the Local Group on different scales affects statistics of the MC analogues, we derive several MW analogue samples that mimic different aspects of the real MW environment: the host halos in the \textbf{P sample} have an M31-sized companion in relative isolation with no other large neighbor; host halos in the \textbf{LGP sample} are a subset of halo pairs from the P sample, but with additional environmental constraints designed to more closely mimic the LG environment; finally, host halos in the  
\textbf{S sample} include all host halos that are not in included in the LGP- and P- samples. These sample definitions and naming convention will be used for both host halos in the MW halo mass range (Section $3.3$) and halo samples in a wide range of masses (Section $4$). Note that for the P  and LGP samples the mass of the M31-like companion is always fixed to the same mass range, even as its MW analogue halo mass is varied within a wider range.

Below the sample definitions are described in more detail.

\begin{figure*}[!htb]
\center
\includegraphics[width=.94\linewidth,angle=0]{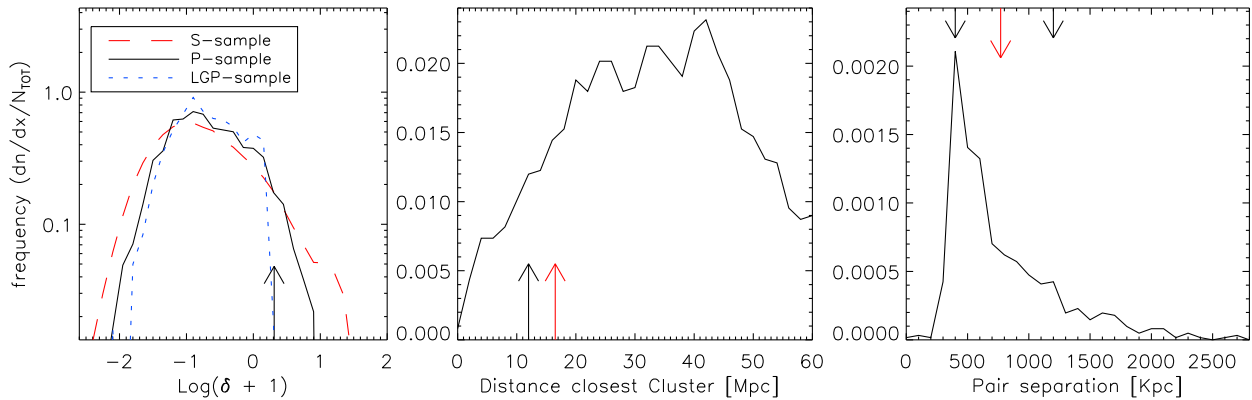}
\caption{
\label{figcuts1} Distribution of the local galaxy overdensity (left panel), distance to the nearest cluster-sized ($M_{200}>1.5 \times 10^{14}$ \msol) halo (center panel), and distance from the pair companion for halos in the P sample (\textit{solid black} line), LGP sample ({\it short-dashed} line), and S-sample ({\it red long-dashed} line). The maximum local overdensity, the minimum distance to the closest cluster, and the minimum pair separation used as additional constraints for the LGP sample of the MW analogues in the LG-like environments are shown by the \textit{black arrows}. The MW is $\sim 16.5 \mpc$ away from the Virgo cluster(middle panel, \textit{red arrow}) and separation between the MW and M31 is $\sim 770 \kpc$(right panel, \textit{red arrow}). }
\end{figure*}

\textbf{The P sample}: the host halos are required to have a companion similar to M31, which we define using the same mass range of  $M_{200c}=0.8 - 2.9 \times 10^{12}$ \msol \citep[e.g.,][]{2010MNRAS.406..264W,2000ApJ...540L...9E}.  The pairs are required to be relatively isolated and not be part of a triplet or a larger group. As a quantitative isolation criterion for the  pair we use the force constraint $F_{i,com}<\kappa F_{12}$, where $F_{i,com}$ is the gravitational force between the pair and any neighbor halo $i$ within a $5 \hmpc$ radius of the pair center-of-mass, and $F_{12}$ is the force between halos in the pair, and $\kappa$ is a constant parameter. The isolation criterion becomes increasingly strict for decreasing values of $\kappa$. The Milky Way and M31 do not have massive neighbors within $5\ \mpc$, and  should thus have $\kappa<0.1$. The actual value of $\kappa$ is, however, uncertain, and  we explore a range of values in the further analysis.
We use $\kappa=0.1$, $0.15$, $0.2$, $0.25$ which results in $205$, $378$, $598$, $810$ LG-like halo pairs, respectively. From the \textit{a posteriori} analysis of satellite properties, results are largely consistent for different $\kappa$ values and we thus use $\kappa=0.2$ in this paper to maximize the statistics of the MW hosts.
The $598$ LG-like halo pairs we found using $\kappa=0.2$ represent $\sim 2\%$ of all halos in the MW mass range, and $\sim 80 \%$ of these pairs are bound under the two-body point mass energy approximation.

\textbf{The LGP sample} of halos is a subset of halos in the P sample with additional constraints on the local and global environment to more closely mimic the environment of the Local Group. Namely, we require that halos in this sample do not have a cluster-sized halo more massive than $M_{200}=1.5 \times 10^{14}$ \msol within 12 Mpc. This mass limit is close to the mass of the Virgo Cluster \msol \citep[e.g.,][and references therein]{2001AA...375..770F,1995MNRAS.274.1093N}. 
For the local environment we compute the galaxy density field using Voronoi tessellation (VT hereafter) on halo positions and masses \citep[similar to the method of][]{2000AA...363L..29S, 2010MNRAS.407.1449G}. The VT partitions the volume into cells, where each cell is associated with a single host halo. The shape and volume of each cell is defined by distribution of halo neighbors. The adaptive local density can be computed using the local cell volume around each halo and the enclosed halo mass, but instead we use also neighbors cell volumes and masses to compute the average, in this way we generate a smooth density field where the typical number of direct neighbors around a halo is $\approx 14$. 
We use only the host halos with mass higher than $M_{200}=1.5 \times 10^{10}$\ \msol for density computation, and define the local overdensity as $\delta=(\rho-\langle\rho\rangle)/\langle\rho\rangle$, where $\langle\rho\rangle$ is the mean of the density distribution. 
The local density of the LG is not well constrained, but the abundance of luminous galaxies within $3-6\ \mpc$ is close to the average density of galaxies in the local universe \citep{2005AJ....129..178K}. Another local environment constraint we include is the distance to the M31-like pair companion. 

Figure \ref{figcuts1} shows the distribution of the local density, distance to the closest Virgo-sized halo, and distance to the M31-sized pair companion for the MW analogue halos in the P sample, and the cuts we impose to define the LGP sample. 
Specifically, we select only halos with the local overdensity smaller $\delta=1.04$ to avoid systems located in the highest density decile. The mean overdensity for the S-sample is located at $\log{(\delta + 1)} \sim 0$, and for the LGP sample it is located $0.5$ dex lower. 

We exclude halos with Virgo-sized neighbors closer than 12 Mpc and require that distance to the pair companion is in the range $0.4 < \Delta r<1.2 \mpc$ to avoid close, possibly merging pairs, but to include pairs with separations similar to the actual distance between the Milky Way and M31. These constraints eliminate about a third of halos from the P sample; the halos in the LGP sample are $\approx 1.3\%$ of the total number of host halos in the MW halo mass range. This indicates that the LG environment of the Milky Way is rather rare for halos of this mass, the fact also indicated by the ``coldness'' of the local velocity field of galaxies \citep[e.g.,][]{klypin_etal03}.

The figure shows that the P sample halos are located in environments with similar overdensity distribution to the S-sample halos, but have a narrower distribution due to the force constraint and the chosen $\kappa=0.2$ value, which eliminates responsible pairs close to other larger halos and single isolated halos.

\section{The MW halo mass estimate}
\label{sec:results}

In the context of the B11 method, in order to estimate the halo mass of the Milky Way from the properties of the MCs, such as circular velocity, velocity and position relative to the center, we explore two aspects which can affect the mass estimate: 1) the environment, in particular whether differences in the environment correspond to  the differences in the subhalo populations of halos, and 2) whether it matters if MCs are treated as two independent dynamical samples or a single tracer (a bound pair sharing a common translational motion of their center of mass). 

\begin{figure}[!bht]
\includegraphics[width=.95\linewidth,angle=0]{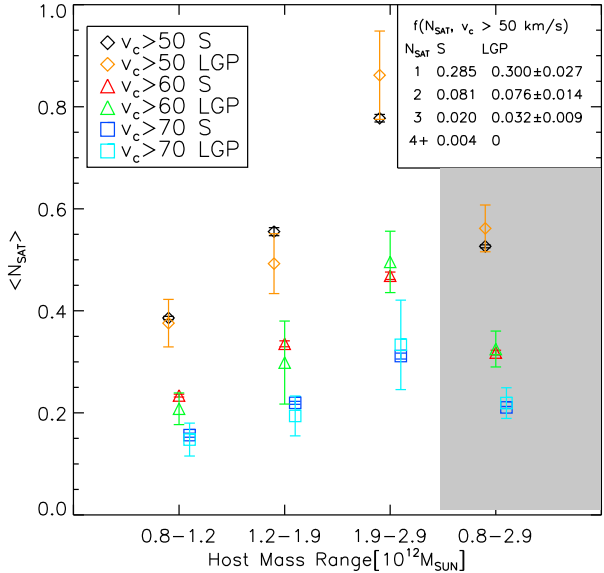}
\caption{
\label{figvc} Average number of satellites in the S and LGP samples for halos in different mass ranges around most likely range of the MW halo mass, and satellite circular velocity ranges defined in legend. The error bars of the mean values are computed using the jackknife method. Top-right legend quotes the fractions of halos with $N_{\rm SAT}$ members for subhalos with $v_{\rm circ}>50$ \kms, and host halo masses in the range $0.8<M_{200}<2.9 \times 10^{12}$ \msol.}
\end{figure}

\subsection{Effect of the environment}

To test for possible effects of the MW environment on subhalo statistics, we 
extract all subhalos with $v_{\rm circ}>50$ \kms, and compute their distance to the host center $r_0$ and the total speed relative to host $s$.
Subhalos very close to the host center ($r_0<20 \hkpc$) are removed to avoid artefacts or resolution-related problems in the halo identification procedure.

Figure \ref{figvc} shows abundance of subhalos with $v_{\rm circ}>50, 60, 70$ \kms in four mass ranges around $10^{12}$ \msol. As expect, the average number of subhalos increases with increasing host halo mass and decreases for increasing subhalo circular velocities. The figure shows that there is no significant difference in the abundances of subhalos in hosts of the S and LGP samples, which means that the Local Group environment does not appreciably affect abundance of the massive subhalos in the MW-sized hosts. 

\begin{figure}[!htb]
\includegraphics[width=.95\linewidth,angle=0]{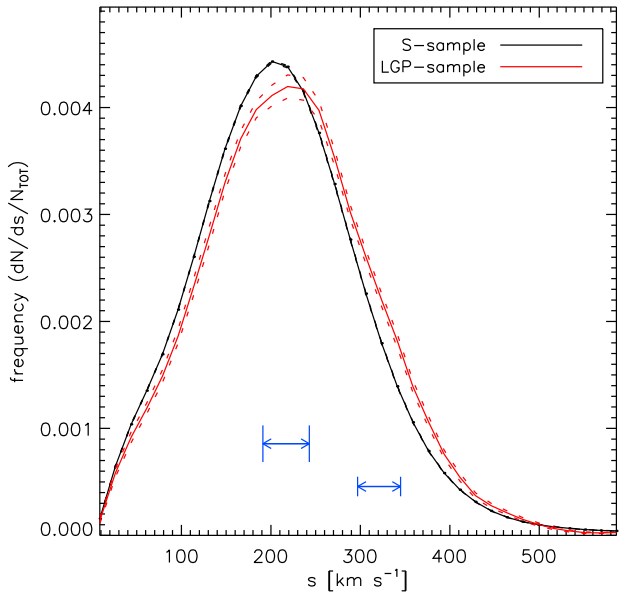}
\caption{
\label{fig2comp} Distributions of the total subhalo velocity in the frame of their host halo for hosts in the S  and LGP samples. Blue arrows indicate the velocities of the SMC and LMC. The uncertainty of the distribution for the LGP sample is shown by the {\it dashed lines} (estimated using the jackknife method).  The environment has small, but statistically significant effect on the distribution of velocities: the fraction of subhalos with the LMC velocity in the LGP sample is $23\pm2\%$ higher than in the S sample. }
\end{figure}

Figure~\ref{fig2comp}  shows the distributions of the total subhalo velocity in the frame of their hosts for halos in the S  and LGP samples.
The fraction of satellites with velocities comparable to that of the LMC is $23\pm2\%$ higher in LGP sample than in the S sample (there is no significant change in the distribution at the SMC speed). 
Therefore, the fraction of satellites matching the velocity of the LMC is somewhat enhanced in the LG-like environments.  This can affect the MW mass estimate because LMC velocity is the main constraint driving the mass likelihood distribution to larger masses (see B11 and Fig.~\ref{figa1} below).

We have tested whether the choice of $\kappa$ parameter used in the isolation criterion (see \S~\ref{sec:LGanalogs}) influences the velocity distribution and have found that the result velocity distribution is almost the same for values of $\kappa$ in the range $0.1<\kappa<0.25$, with a weak trend towards higher fraction of subhalos in the LGP sample matching the velocities of the LMC and SMC for lower $\kappa$ values. 
The magnitude of the trend, however, is comparable to the uncertainties in the velocity distribution. 

\begin{figure}[!htb]
\includegraphics[width=.95\linewidth,angle=0]{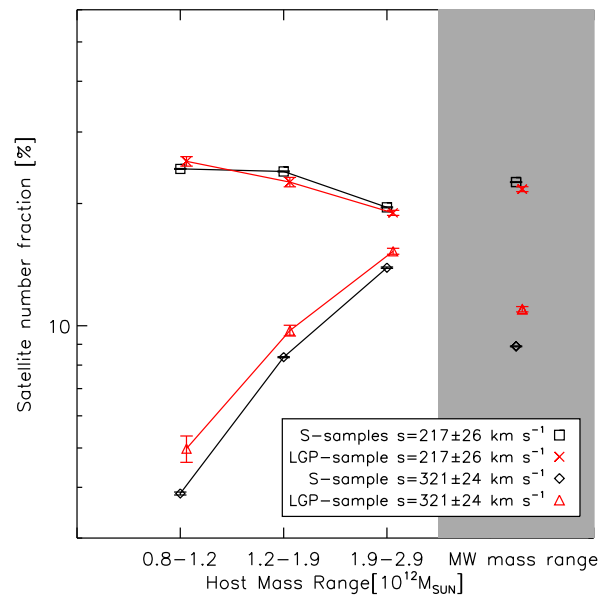}
\caption{
\label{fig2compb} Fraction of subhalos matching the velocity of the SMC and LMC in the S ({\it black symbols and lines}) and LGP ({\it red symbols and lines}) samples for three different halo mass ranges around $M_{200}=10^{12} \msol$ (the first three connected bins) and the total mass range used to define the MW halo analogues ({\it the gray shaded} region).}
\end{figure}

In figure \ref{fig2compb} we show the fraction of subhalos with total velocities  relative to the host center similar to those of the LMC and SMC for host halos in different mass ranges around $M_{200}=10^{12} \msol$. There is a small enhancement in the frequency at the LMC velocity, but no difference at the  SMC velocity. 
In the LGP sample the fraction of host halos with subhalos matching the LMC velocity is larger by $29\pm10\%$, $16\pm4\%$, and $10\pm2\%$ than in the S sample for $0.8-1.2$, $1.2-1.9$, and $1.9-2.9 \times 10^{12} \msol$ mass ranges respectively.

\begin{figure}[ht]
\includegraphics[width=.95\linewidth,angle=0]{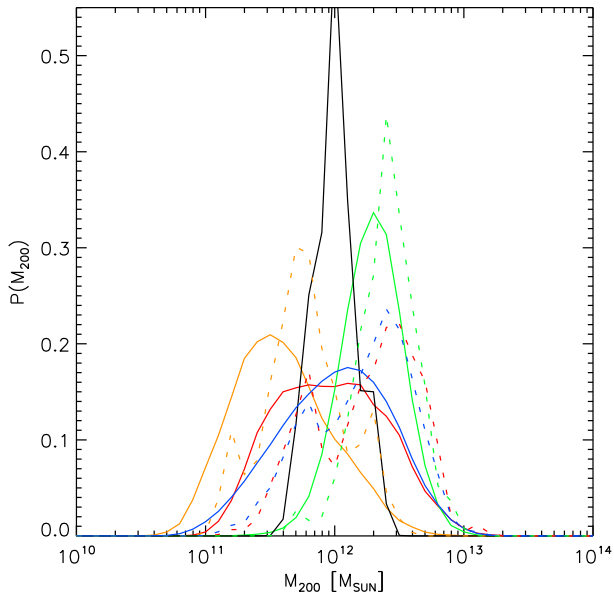}
\caption{
\label{figa1} Likelihood distribution for the MW halo mass ({\it black} line), $M_{200}$,  in the S sample based on the properties of the LMC and SMC allowing the full halo mass range instead of the range $M_{200}=0.8-2.9\times10^{12}M_{\odot}$. Lines of  different colors show likelihoods when only one constraint is used:  exactly 2 subhalos with $v_c>50 km/s$ ({\it blue} line), circular velocities of the MCs ({\it red} line), radial positions of the LMC and SMC ({\it orange} line), total velocity relative to the host center of the LMC and SMC ({\it green} line). {\it Dotted lines} are corresponding distributions for the LGP sample using the same color key, but we omit the distribution using all constraints together in this case due lack of matching systems.
The MW mass estimate in the S sample is $\log(M_{200}/\msol)=12.06^{+0.31}_{-0.19}$ ($90\%$ confidence interval).}
\end{figure}

We present the likelihood distribution for the MW halo mass from the constraints from the Magellanic Cloud properties in Figure~\ref{figa1}. To compute the likelihood we used the S and LGP sample constructed without any restriction on the $M_{200}$ mass of halos. In addition to the likelihood for all of the constraints combined for S-sample only, we also show likelihoods resulting from using only one of the constraints for both S and LGP samples. In particular, we present the likelihood for the constraint that host halos have exactly two subhalos with $v_{\rm circ}>50$ \kms with all other constraints (circular velocity, radial distance, and total velocity) applied after this condition to compare to the results of B11.

We chose not to enforce the two satellite condition in our estimate of the likelihood, instead we allow the host halo to have any number of satellites to increase the sample size, but we will consider only the two largest ones. The mass estimates obtained with and without enforcing the two satellite condition are presented in Table \ref{table1}. 
The final mass estimate using all of the constraints together is only computed for the S sample, in which there are $40$ MC-like satellite pairs. There are only 3 such satellite pairs in the LGP sample, and therefore we do not attempt to derive the total likelihood for this sample. However, given that the likelihood distributions for individual property constraints are similar for the S and LGP samples, we expect that the total likelihood is similar for the combined constraints as well. 

The MW halo mass estimate for the S sample is $\log(M_{200}/\msol)=12.06^{+0.31}_{-0.19}$ ($90\%$ confidence interval), which is in general agreement with the B11 result in the central value. We present results for different samples and constraints in Table \ref{table1}, where we also include both the $68\%$ and $90\%$ confidence interval errors. Due to significant deviations of the likelihood from the log-normal form in the tails, the $90\%$ errors are significantly larger than the $68\%$ ones and we therefore chose to quote the $90\%$ errors. 

Note that there is a key methodological difference in the way  the likelihood distribution was evaluated in B11 and in our analysis. In our mass estimate we use all the constraints together simply as the distribution of halo properties that satisfy the constraints. This way any correlations between properties, such as expected correlation between radial distance to the host center and total velocity are taken into account automatically. B11, on the other hand, assume that the probability distributions for each constraint are independent (see their eq. 3). If we multiply the probability distributions for radial, speed and circular velocity constraints we get an estimate of the MW mass which should be similar to the method of B11. For the S sample this gives $\log(M_{200}/\msol)=12.24^{+0.25}_{-0.25}$ and for the LGP sample $\log(M_{200}/\msol)=12.44^{+0.24}_{-0.22}$. Thus, treating constraints as independent results in a small overestimate of the mass.

\subsection{Dependence on the SMC and LMC constraints}

The second key assumption of the B11 analysis is that the properties of the two Magellanic Clouds are not correlated. However, the MCs are likely a bound pair, and their velocities and radial positions can thus be expected to be correlated. We have tested whether this assumption affects the MW mass estimate by comparing results in the following two cases: 1) properties of the MCs are treated as independent and we compare mass likelihood using both MCs, only the LMC, or only the SMC; 2) the MCs are considered to be a bound pair and we compute the likelihood using average properties of the pair, rather than properties of the two MCs independently. We explore the latter scenario in the next subsection.

For the case 1 we compute the mass likelihood for the LMC and SMC independently and find $\log(M_{200}/\msol)=12.18^{+0.33}_{-0.23}$ using the LMC only, consistent with the result using both MCs discussed above. For the case when we use velocity of the SMC only, the mass is not well constrained as the likelihood extends to considerably lower masses. Thus, the MW mass estimate is dominated by the properties of the LMC, as is expected since it is more massive and has a higher velocity. 

The fact that the main mass constraint comes from the satellite with the largest velocity is generic. If one considers distribution of absolute magnitude of satellite velocities as a function of the radial distance to the host center, at any given $r$ the distribution is broad but has a sharp cut off at the velocity close to the escape velocity of the host. It is this sharp cut off that constrains the mass, and the constraint is due to the satellite with the largest absolute velocity at a given radius. This fact was recently used to constrain the halo mass of the Milky Way with Leo I satellite by \citet{boylan_kolchin_etal12}.

In principle, we can include additional properties of the LMC and SMC, such as as their separation or relative velocity \citep[$\approx 23 \kpc$ and $128 \pm 32$ \kms, respectively][]{2006ApJ...652.1213K,kal2012}, in the derivation of the mass constraint. However, this is difficult in practice because very few host halos in the Bolshoi simulation match all of the properties of the LMC and SMC. We examine the incidence of the MC-like close pairs of satellites in the next section. 

Here we adopt a different approach, in which we assume that the LMC and SMC are a bound pair and can be considered as a single subhalo. We then consider the pair as a single velocity tracer and use the velocity of the center-of-mass of the pair as a constraint. We calculate the center-of-mass velocity assuming that the MCs move in the same direction and neglecting peculiar velocity around the center-of-mass: $\|\vec{s}_{\rm cm}\|=\|\vec{s}_{1}M_{1}+\vec{s}_{2}M_{2}\|/(M_{1}+M_{2})=298\pm52$ \kms. We also treat the pair as a single subhalo with the circular velocity of 
$v_{\rm circ}=85.4^{+16}_{-10}$ \kms. The masses are evaluated numerically from the $M_{200}-v_{\rm circ}$ relation for subhalos in the entire Bolshoi simulation. 
The asymmetry in the error range is because we include larger values for the LMC $v_{\rm circ}$ \citep{2011ApJ...737...29O}.
Finally, we adopt the radial distance of the center-of-mass of $r_{\rm cm}=(r_1M_1+r_2M_2)/(M_1+M_2)=54\pm5$ \kpc.

\begin{figure}[!htb]
\includegraphics[width=.95\linewidth,angle=0]{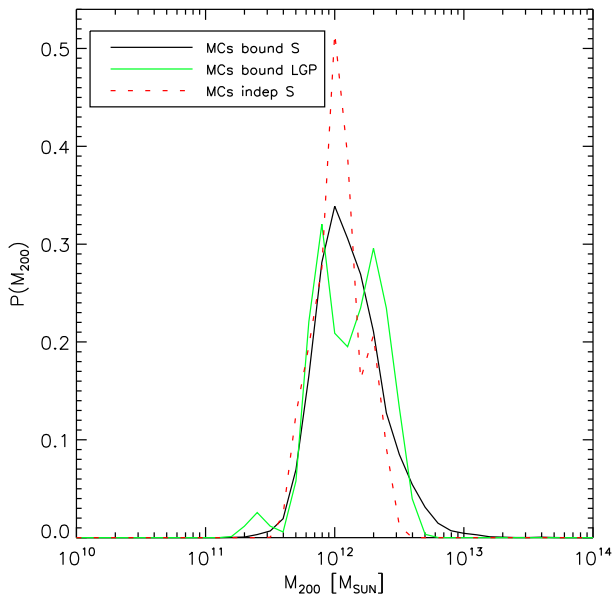}
\caption{
\label{figlmcsmc} Likelihood distribution of the Milky Way halo resulting from the SMC and LMC being considered a bound system corresponding to a single subhalo (with parameters quoted in the text). The distribution  for the S sample is shown by the {\it black solid} line,  and for the LGP sample by the {\it green solid} line. The corresponding halo mass constraints are $\log(M_{200}/\msol)=12.13^{+0.36}_{-0.25}$ and $\log(M_{200}/\msol)=12.17^{+0.30}_{-0.34}$ (errors correspond to the $90\%$ confidence level) for the S and LGP samples, respectively. Mass estimate for the S sample from Fig.\ 1 derived when properties of both MCs are used separately is shown for comparison by the {\it dotted red} line.}
\end{figure}

Figure \ref{figlmcsmc} shows the likelihood distribution for the MW halo mass estimated with such constraint for the S and LGP samples. We also include the mass estimate from figure \ref{figa1} for comparison. The halo mass in this case is constrained to be  $\log(M_{200}/\msol)=12.13^{+0.36}_{-0.25}$ ($90\%$ confidence interval) in the S sample and $\log(M_{200})=12.17^{+0.30}_{-0.34}$ in the LGP sample. We do not find any significant variation with environment in this case, and the mass estimate errors are consistent with the case when we constrain the mass using both MCs as independent tracers. The only difference is that the use of the MCs as a single tracer eliminates the tail of the likelihood towards small masses ($M_{200}\sim 10^{11}\ \rm M_{\odot}$), which is due to the fact that subhalos with circular velocities as high as $v_{\rm circ}=85$ \kms  are highly unlikely in such small host halos.   As noted above, the lower velocity SMC does not influence the constraint  due to the fact that the bulk of the constraint is due to the LMC that has the largest velocity. The addition of the SMC either as independent tracer or as a second object in a pair to get the average center-of-mass values does not influence the constraint appreciably. 

The summary of all the mass constraints is presented in Table~\ref{table1}.
In the left half of the table we show the constraints obtained using host halos with any number of satellites with $V_{circ}>50$ \kms, while in the right half we show results obtained with the requirement that host halos host exactly two satellites with $V_{circ}>50$ \kms, as used in the study of B11. The
first column lists the constraints for the S sample, for the case where we multiply the different likelihood distributions for each constraint instead of computing the mass likelihood directly (S-multi and LGP-multi), the case when only properties of the LMC are used in the constraint, and the case where we assume that the SMC and LMC are a bound pair and correspond to a single subhalo with properties given by the mass-weighted average of the SMC and LMC properties for the S and LGP samples (S-bound and LGP-bound).
Columns $2-4$ and $5-7$ show the mass constraints and the errors corresponding to the $68\%$ and $90\%$ confidence intervals. $M_{200}$ masses can be converted to $M_{vir}$ using the conversion factor of
$M_{vir}/M_{200}=1.21$, computed directly the Bolshoi simulation for the halos of this mass range.

\begin{table*}[ht]
\caption{\label{table1} Constraints on the halo mass of the Milky Way using different host samples and properties of the LMC and SMC }
\centering
\begin{tabular}{lcccccc}
\hline\hline\\
 & Any number subs & $v_{\rm circ}>50$ \kms & &  Exactly two subs & $v_{\rm circ}>50$ \kms & \\
Sample & $\log(M_{200}/M_{\odot})$ & $68\%$ c.i. & $90\%$ c.i. & $\log(M_{200}/M_{\odot})$ & $68\%$ c.i. & $90\%$ c.i\\[2mm]
\hline\\
S-sample   &  12.06 &  +0.08 -0.05 & +0.31 -0.19 &  12.03 & +0.06 -0.02 & +0.34 -0.17 \\
S- multi   &  12.24 &  +0.13 -0.20 & +0.25 -0.25 &  12.15 & +0.15 -0.05 & +0.24 -0.30 \\
LGP- multi &  12.44 &  +0.07 -0.13 & +0.24 -0.22 &  12.26 & +0.14 -0.06 & +0.23 -0.19 \\
LMC only   &  12.18 &  +0.10 -0.10 & +0.33 -0.23 &  12.17 & +0.10 -0.11 & +0.31 -0.24 \\
S-bound    &  12.13 &  +0.13 -0.11 & +0.36 -0.25 &  12.09 & +0.11 -0.09 & +0.28 -0.23 \\
LGP-bound  &  12.17 &  +0.16 -0.19 & +0.30 -0.34 &  11.99 & +0.16 -0.17 & +0.47 -0.19 \\
\\
\hline
\end{tabular}\\[2mm]
Notes: columns $2-4$ show the constraints for the case when host halos are allowed to have any number of satellites with $v_{\rm circ}>50$ \kms, while columns $5-7$ show the corresponding constraints for the case when host halos are restricted to have exactly two satellites with $v_{\rm circ}>50$ \kms. We present the mass errors corresponding to both the $68\%$ and $90\%$ confidence intervals from the derived likelihood distributions. 
\end{table*}

\subsection{Close satellite pairs}
\label{sec:closepairs}
\begin{figure*}[ht]
\center
\includegraphics[width=.9\linewidth,angle=0]{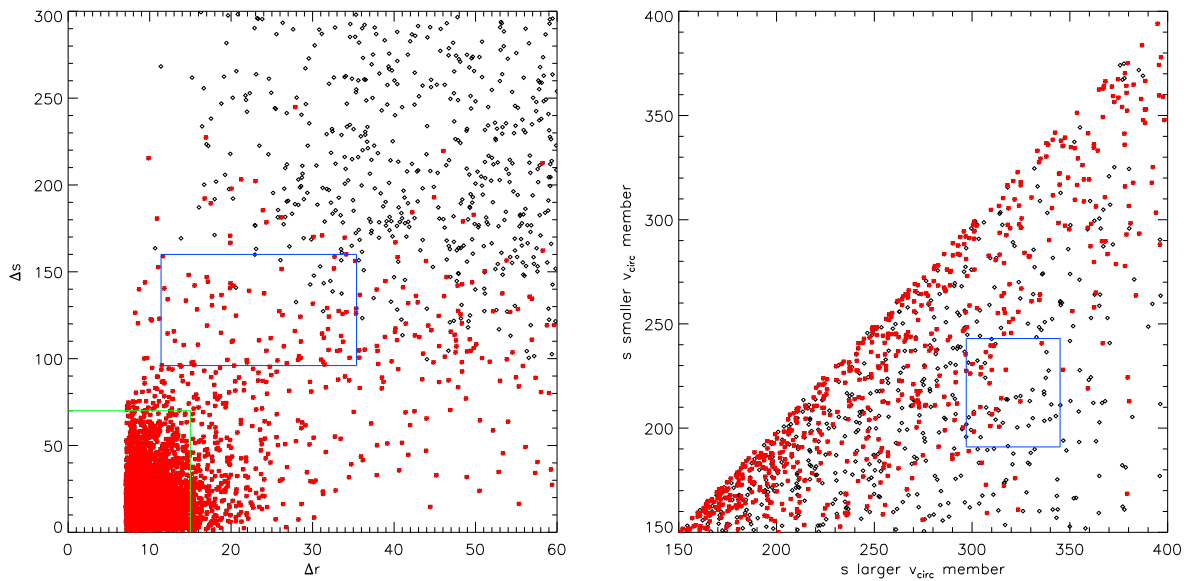}
\caption{
\label{figallcp} Properties of close subhalo pairs in MW mass halos. Left panel: pair separation and relative velocity; right panel: total speed relative to the host center for both pair members(horizontal axis for the member with larger circular velocity). Red points indicate subhalo pairs which are bound if we assume a two-body system. Blue rectangles show the MCs constraints ($\Delta r<60$ \kpc, $\Delta s<300$ \kms, $150<s<400$ \kms, $44<r_0<66$ \kpc, and $v_{\rm circ}>50$ \kms). Green box in the left panel indicates very close pairs $\Delta r<15$ \kpc, $\Delta s<70$ \kms which contain a large fraction of fake pairs (see text for details). }
\end{figure*}

Satellite pairs are quite rare: we find that only $\approx 2\%$ of the MW-sized halos have a pair of subhalos with $v_{\rm circ}>50$km/s, separation $<40 \kpc$, and relative velocity $<160$ \kms. This fraction drops to $\sim 1/30000$ for pairs more closely resembling the SMC--LMC pair.

In this section we explore the incidence of the population of close subhalo pairs and test whether their properties are distinct from the properties of the overall subhalo population. It is important to clarify these issues, because velocities and other properties of the SMC and LMC may be influenced by their mutual interaction  \citep[e.g.,][]{2008ApJ...686L..61D,2011ApJ...742..110N,2011MNRAS.414.1560B}. To select a sample of close pairs, we consider subhalos with $v_{\rm circ}>50$ \kms within the MW-sized halos that have distance to the host center, mutual separation and absolute velocity similar to those of the MCs: $44<r_0<66$ \kpc, $\Delta r<60$ \kpc, $150<s<400$ \kms, and $\Delta s<300$ \kms. We also estimate if the pairs are bound by computing the total energy assuming a two body system of two point masses with masses of the SMC and LMC as derived from the $v_{\rm circ}-M$ relation in the simulation. Such approximation should be viewed as a rough approximation for the fraction of bound systems in a given sample, but can be used as a reference.

In figure \ref{figallcp} we show the separation and relative velocities of the identified pairs (left panel) and velocity in the frame of the host halo of both pair members (right panel), with the $x$-axis showing velocity of the pair member with the larger circular velocity. Red points indicate pairs that are bound according to our energy estimate. The figure shows that only a small fraction of pairs has properties similar to those of the SMC--LMC pair $\Delta r_{\rm MC}=23.4\pm10$ \kpc, and $\Delta s_{\rm MC} =128\pm32$ \kms (shown by the blue boxes in the two panels). 
The adopted range of separation of $23.4\pm10$ \kpc is considerably larger than the actual observational error, which is of the same order as the error in the galactocentric distance: $\approx 2$ \kpc \citep{2006ApJ...652.1213K}, but it allows us to select MC-like pairs without imposing a prohibitively restrictive constraint on the subhalo pair configuration.
Only $\approx 60$ satellite pairs out of $1140$ outside the green region have separations and relative velocities similar to those of the SMC and LMC, $\sim 90\%$ of which are bound according to our criterion. 

The left panel of Fig.~\ref{figallcp} shows a rather large number of pairs clustered at separation $\lesssim 15 \kpc$ and $\Delta s\lesssim 70$~kpc. Most of these pairs appear to be artefacts due to failures of the halo finder for subhalo candidates close to the resolution limit of the simulation.  We have visually inspected the dark matter density and velocity fields around a representative subset of these very close pairs and found that at separation $10-20 \kpc$ $\approx 20\%$ subhalo pairs do not have corresponding distinct peaks in dark matter density field or coherently moving clumps of dark matter (i.e., subhalos are fake). Some of the pairs appear to be due to  misidentification, in which a subhalo undergoing a tidal disruption is identified as two subhalos with circular velocities close to the resolution limit. On the other hand, we find that for pairs with separation $>20 \kpc$, almost all pairs have two clear distinct density peaks in the dark matter distribution and corresponding coherent velocity streams in the velocity field. 

\begin{figure}[ht]
\includegraphics[width=.95\linewidth,angle=0]{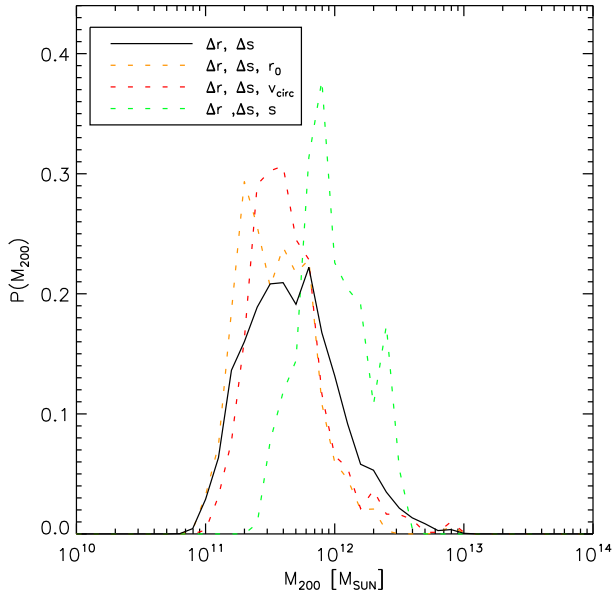}
\caption{
\label{figpmclose} Likelihood distribution of the Milky Way halo mass using additional constraints that subhalo pairs have separation and velocity difference similar to those of the SMC--LMC pair: $\Delta r_{MC}=23.4\pm10$ \kpc, and $\Delta s_{MC} =128\pm32$ \kms ({\it black solid} line). Likelihood distributions for subhalo pairs matching only the radial distance ({\it orange} line), circular velocity ({\it red} line), or relative velocity with respect to the host halo ({\it green} line) are shown by the dotted lines. Less than $21\%$ of close pairs are in halos of $M_{200}>10^{12} \msol$.   }
\end{figure}

If we add the constraints that the pair of subhalos in a host halos must have separation $\Delta r_{MC}=23.4\pm10$ \kpc, and $\Delta s_{MC} =128\pm32$ \kms, the resulting likelihood for the Milky Way host halo mass is shown  in figure \ref{figpmclose}. Due to small number of close subhalo pairs satisfying the constraints, the mass likelihood distribution can be computed using only one of the MCs constraints at a time together with the separation and velocity difference constraints.  
We find that $<21\%$ of close pairs are in halos of $M_{200}>10^{12} \msol$.\footnote{This fraction is for the host halos restricted to have exactly two satellites with $V_{circ}>50$ \kms. If we allow any number of satellites, this fraction increases to $26\%$.} The addition of the constraint on the pair separation and relative velocity thus pushes the MW halo mass constraint to considerably lower masses, as compared to the constraint with only the properties used by B11. This illustrates that the actual constraint depends quite sensitively on which properties of the SMC--LMC system are chosen for the analysis, as is also clear from the large differences between mass likelihood distributions for the individual properties shown in Fig.~\ref{figa1}. 

\section{Discussion and conclusions}
\label{sec:conclusions}

In this study we have explored whether the environment of the Milky Way affects properties of its two most massive satellites, the SMC and LMC. These satellites are rather rare around galaxies with the Milky Way luminosity and have rather high velocities. As argued by \citet{2011ApJ...743...40B}, the properties of the SMC and LMC can be used for a useful independent estimate of the Milky Way virial mass. However, it would be straightforward to interpret such estimate only if the Magellanic Clouds are not very unusual and their properties are not biased by the environment. 

To test for the effects of the environment, we compare velocity distributions for a sample of MC-sized subhalos around Milky Way hosts selected simply by mass and a sample of such halos selected with additional restrictions on the distance to the nearest cluster and local galaxy density, designed to mimic the environment of the Local Group. We find that the velocity distribution of satellites in the latter sample is shifted somewhat to larger velocities: in particular, the fraction of satellites with the LMC speed in the LGP sample is $23\pm2\%$ larger than in the S sample. Thus, the Local Group environment enhances the probability of high satellite velocities, although the effect is mild. 

We compute the likelihood distribution for the Milky Way halo mass using the method similar to that of \citet{2011ApJ...743...40B}, and explore how this distribution depends on different properties of the SMC and LMC used as constraints. We find no significant effect of the environment on the mass estimate. We also find that the treatment of different MC properties as mutually independent does not bias the mass constraint. 

Using properties of the SMC and LMC from the recent study by \citet{kal2012}, we derive constraint on the MW halo mass of $\log(M_{200}/\msol)=12.06^{+0.31}_{-0.19}$ ($90\%$ confidence interval) for the S sample of halos selected without any environment restrictions. The mass constraint we derive is similar to that of B11, even though the updated values of MC properties, such as velocities and their errors, are quite different from the values used by B11. The mass constraint is broadly consistent with other recent estimates of the Milky Way halo mass. 

The method of B11 does therefore appear to be a robust way to measure the MW halo mass. A subtle issue, however, is that if some of the properties of the LMC or SMC used for the constraint are rare for the systems of the MW mass, the interpretation of the mass likelihood is not straightforward. The Milky Way in this case may be located in the tail of the distribution and interpretation of the peak of the likelihood as most likely mass of the MW halo is not correct.   It remains to be seen how rare particular properties of the LMC and SMC are. We do know, for example, that the probability of having two satellites of the SMC and LMC luminosity is by itself quite rare in the hosts of the MW luminosity \citep[e.g.,][]{liu_etal11,2012MNRAS.424.1448R}.

LMC and SMC are not just two unrelated luminous satellites: they are a close pair.  
We show in this study that close satellite pairs are quite rare: pairs with masses and separations similar to those of the LMC and SMC are found only in one out of $\approx 30000$ MW-sized halos. Observations also indicate that such close pairs are very rare \citep{2011MNRAS.411..495J}. We find that satellites in most close pairs with properties similar to the MCs are likely to be bound to each other. Interestingly, the halo mass likelihood distribution for host halos constrained to have MC-like close pairs of subhalos ($\Delta r_{\rm MC}=23.4\pm10$ \kpc, and $\Delta s_{\rm MC} =128\pm32$ \kms) is quite different from the global likelihood from which the MW halo mass constraint discussed above was derived. In particular, less than $21\%$ of host halos with $M_{200}>10^{12} \msol$ host MC-like close pairs. Taking into account the close separation of the MCs in the B11 method results in the shift of the MW halo mass estimate to smaller masses with the peak shifting approximately by a factor of two (see Fig.~\ref{figpmclose}). 
The reason for this shift is the fact that in smaller halos it is more likely to get a pair with small separation and relatively small velocity difference by chance (not necessarily bound), mainly because velocity dispersion is smaller in smaller mass halos. 

This example clearly shows that a great care should be taken in choosing which of the satellite properties are used for the MW mass constraint.

\acknowledgments 
This work was supported by NSF via grant OCI-0904482. 
AK was in addition supported in part by NSF
grants AST-0807444 and by the
Kavli Institute for Cosmological Physics at the University
of Chicago through the NSF grant PHY-0551142 and PHY-
1125897 and an endowment from the Kavli Foundation. We have made extensive use of the NASA Astrophysics Data System
and {\tt arXiv.org} preprint server.


\bibliography{satbib}

\end{document}